# Effect of superconducting correlation on the localization of quasiparticles in low dimensions

Tao Xiang

*Research Center in Superconductivity, University of Cambridge, Madingley Road, Cambridge CB3 0HE, United Kingdom*

(March 28, 1995)

Localization lengths of superconducting quasiparticles $\lambda_s$ are evaluated and compared with the corresponding normal state values $\lambda_n$ in one and two dimensional lattices. The effect of superconducting correlation on the localization of quasiparticles is generally stronger in an off-site pairing state than in an on-site pairing state. The modification of superconducting correlation to $\lambda$ is strongly correlated with the density of states (DOS) of superconducting quasiparticles. $\lambda_s$ drops within the energy gap but is largely enhanced around energies where DOS peaks appear. For a gapless pairing state in 1D or a d-wave pairing state in 2D, $\lambda_s/\lambda_n$ at the Fermi energy $E_F$ is of order 1 and determined purely by the value of gap parameter not by the random potential. For the d-wave pairing state, the localization effect is largely weakened compared with the corresponding normal state and quasiparticles with energies close to $E_F$ are more strongly localized than other low energy quasiparticles.

There has been a renewal of interest in the study of the localization effect of quasiparticles in superconducting states. Recently in explaining the thermally activated behavior of the microwave conductivity in high $T_c$ superconductors, Lee has proposed that a conductivity mobility gap exists in the $d_{x^2-y^2}$-wave pairing state [1]. Since a finite density of states (DOS) exists at the Fermi level in a disorder d-wave superconducting state [2] he argued that this system should formally resemble a metal with a finite Fermi surface where the low energy quasiparticles are strongly localized. More recently Hatsugai and Lee [3] have numerically studied the effect of impurities on Dirac fermions on a square lattice and found that low energy states are strongly localized compared with high energy states. Their result lends support to Lee's argument as the spectrum of the Dirac fermion is similar to the d-wave superconductor. However, the superconducting correlation is not included in this Dirac fermion model.

The effect of localization in superconductors has actually been studied for many years, with the emphases on either the change in $T_c$ or the effective electron-electron (-phonon) interactions [4]. The weak localization of quasiparticles in the s-wave superconductor was first studied by Ma and Lee [5] in 1985. Within a mean-field approximation, they found that superconductivity persists below the mobility edge and the quasiparticle excitations of the superconductors formed from localized states are also localized. Purely off-diagonal disorder (i.e. a spatial fluctuated gap function) can also localize superconducting quasiparticles. This problem has been addressed by Lambert et al [6].

In this paper, we study the effect of superconducting correlation on the localization of quasiparticles by evaluating the localization lengths of quasiparticles in one and two dimensions. The model used here is the weak-coupling BCS Hamiltonian with disorder which is defined by

$$H = t \sum_{<\mathbf{rr'}>\sigma} c^\dagger_{\mathbf{r}\sigma} c_{\mathbf{r}'\sigma} + \sum_{\mathbf{r}\tau}(\Delta_{\mathbf{r}\tau} c^\dagger_{\mathbf{r}\uparrow} c^\dagger_{\mathbf{r}+\tau\downarrow} + h.c.)$$
$$+ \sum_{\mathbf{r}\sigma}(V_\mathbf{r} - \mu)c^\dagger_{\mathbf{r}\sigma} c_{\mathbf{r}\sigma}, \qquad (1)$$

where $<\mathbf{rr'}>$ denotes nearest neighbors and $\mu$ is the chemical potential. $V_\mathbf{r}$ is a white noise potential which distributes uniformly between $-V$ and $V$. $\Delta_{\mathbf{r}\tau}$ is the superconducting order parameter which should be determined self-consistently from the Bogoliubov-de Gennes equations. Without disorder, $\Delta_{\mathbf{r}\tau}$ is translationally invariant with a particular symmetry with respect to $\tau$. With disorder the self-consistent gap function $\Delta_{\mathbf{r}\tau}$ is space dependent. To solve the Bogoliubov-de Gennes equations in this case is rather difficult. For simplicity we shall ignore the fluctuation of $\Delta_{\mathbf{r}\tau}$ in space. We assume $\Delta_{\mathbf{r}\tau} = \Delta_\tau$ and take $\Delta_\tau$ as a free gap parameter. This approximation neglects the contribution of off-diagonal disorder to the localization of quasiparticles. But as we have shown recently [7], the error resulted from the approximation is small provided $\Delta_\tau$ is given by the average of the self-consistent energy gap $\Delta_{\mathbf{r}\tau}$. Moreover, in the weak disorder limit ($V \longrightarrow 0$) the fluctuation of $\Delta_{\mathbf{r}\tau}$ in space is negligibly small. Results obtained in this limit should be rigorous.

In the absence of superconducting correlations, $H$ is the ordinary disorder Anderson Hamiltonian. It is well known that all states are localized in one and two dimensions in this model according to the standard scaling theory of localization [8]. The presence of superconducting correlation may modify the properties of localization of electrons in two aspects. First, the presence of superconducting correlations induces long range correlations between electrons which may enhance the coherent motion of electrons. Second, the presence of superconducting correlations opens a gap in low-lying excitations. The opening of this gap suppresses the low energy DOS (it is zero within the gap for a clean s-wave superconductor) which may reduce the mobility of electrons and make low energy quasiparticles more localized. On the other hand the DOS right above the gap edges ($E \sim \Delta$) diverges in the clean limit. With disorder the DOS is no longer divergent, but still very large at the gap edge which may enhance the mobility of electrons.

The Hamiltonian (1) is bilinear in fermion operators. For the convenience in the discussion below, we perform a particle-hole transformation for the down-spin electrons, $c_{\mathbf{r}\downarrow} \longleftrightarrow c^{\dagger}_{\mathbf{r}\downarrow}$, and re-express (1) as

$$H' = t \sum_{<\mathbf{rr'}>} (c^{\dagger}_{\mathbf{r}\uparrow}c_{\mathbf{r'}\uparrow} - c^{\dagger}_{\mathbf{r}\downarrow}c_{\mathbf{r'}\downarrow}) + \sum_{\mathbf{r}\tau}(\Delta_{\mathbf{r}\tau}c^{\dagger}_{\mathbf{r}\uparrow}c_{\mathbf{r}+\tau\downarrow} + h.c.)$$
$$+ \sum_{\mathbf{r}}(V_{\mathbf{r}} - \mu)(c^{\dagger}_{\mathbf{r}\uparrow}c_{\mathbf{r}\uparrow} - c^{\dagger}_{\mathbf{r}\downarrow}c_{\mathbf{r}\downarrow}) + const. \quad (2)$$

$H'$ has the usual tight binding model form, but the hopping constant, the chemical potential, and the impurity potential have opposite signs for the up spin electrons and the down spin electrons. The superconducting pairing term in (1) now becomes the spin flipping term in (2).

We evaluate the localization length using an iterative method applied to very long strips combined with a finite size scaling analysis [9]. The localization length on a long strip is defined from the one particle Green's function G(E) according to the relation

$$\lambda^{-1}(M, E) = -\lim_{N \to \infty} \frac{1}{2N} \sum_{ij\sigma\sigma'prime}^{M} |G_{1i\sigma, Nj\sigma'}(E)|^2, \quad (3)$$

where N and M are respectively the length and the width of the strip. M=1 is the true one-dimensional case. In our calculation, N is generally terminated when N is more than 100000 in 1D or when N is 1000 times larger than the localization length $\lambda(M, E)$ in 2D. In averaging $\lambda$ over the disorder, we have taken 200 random configurations of disorder in 1D and 10 random configurations of disorder in 2D. The relative error estimated from the deviation of $\lambda$ around the mean value is less than 3%.

We first consider the 1D case. An on-site pairing state $\Delta_\tau = \Delta \delta_{\tau,0}$ and an off-site pairing state $\Delta_\tau = \Delta(\delta_{\tau,\hat{x}} + \delta_{\tau,-\hat{x}})$ at or away from half filling will be investigated. The calculation in 1D is simple. It nevertheless reveals some intrinsic properties of localization effects in superconducting states, which may help us understand the effect of superconducting correlation on the Anderson localization in high dimensions. At half filling ($\mu = 0$) the off-site pairing state has a gapless excitation spectrum with a finite DOS at the Fermi level. This special case provides a chance for understanding the localization effect in gapless superconductors from 1D.

Fig. 1 shows $\lambda_s/\lambda_n$ (the ratio of the localization length between a superconducting state and the corresponding normal state) as functions of energy E for the on-site pairing state with $\mu = 0$ and the off-site pairing state with $\mu = 0$ and $-0.5$ in 1D. For comparison, the corresponding DOS is also shown in Fig. 1. Essentially three features are revealed by the figure: 1. $\lambda_s/\lambda_n$ and DOS are highly correlated. For the pairing states with finite energy gaps, (b) and (c) in the figure, $\lambda_s/\lambda_n$ drops very fast within the energy gap. On the other hand, $\lambda_s/\lambda_n$ shows a peak around an energy where a DOS peak appears. The higher the peak in the DOS, the higher the peak in $\lambda_s/\lambda_n$ [10]. The drop of $\lambda_s/\lambda_n$ around $E \sim 1.5$ in Fig 1b is a result of the sudden drop in the DOS around that energy. 2. For the off-site pairing state $\lambda_s/\lambda_n$ is enlarged in nearly the whole energy region at half filling ($\mu = 0$). This enhancement is determined only by $\Delta$, independent of the random potential V, when E is not in the region where singular behaviors in the DOS appear or very close to $E_F$ (=0 in the figure). For the on-site pairing state the enhancement in $\lambda_s/\lambda_n$ is very small outside the energy gap region, very different from the off-site pairing state. 3. For the off-site pairing state at half filling, $\lambda_s/\lambda_n$ drops near $E_F$, but only in a narrow energy region which is potential dependent (see the inset of Fig. 1a). Low energy states are therefore more localized than high energy states in this gapless pairing state. The drop of $\lambda_s/\lambda_n$ around $E_F$ is not due to the effect of the DOS, since in this case the DOS is finite and almost unchanged by disorder. Exactly at $E_F$, $\lambda_s/\lambda_n$ is only a function of $\Delta$, independent of V as shown in Fig. 2. This might be a common property of gapless superconductors. As will be shown later, this universal behavior of $\lambda_s/\lambda_n$ at $E_F$ appears also in the d-wave pairing state in 2D.

The difference in the localization effect between the on-site and the off-site pairing states is in fact intrinsic. To understand this point, let us consider the Hamiltonian (2). As mentioned previously, after the particle-hole transformation, the pairing term becomes the spin flipping term and the impurity potential has opposite sign for up-spin electrons and down-spin electrons. This sign difference is not important in the absence of the spin flipping term where up-spin electrons and down-spin electrons are decoupled. However, in the presence of the spin flipping term, this sign change has a significant consequence. Let us consider, for example, a up-spin electron hopping from site **i**, where the potential is negative for up-spin electrons, to one of its nearest neighbors, site **j**, where the potential could be either positive or negative for up-spin electrons. For the nearest-neighbor pairing state, this up-spin electron can hop from site **i** to site **j** in two ways. One is to hop without flipping spin (i.e. hopping through the $t$-term), the other is to hop and at the same time flip its spin (i.e. hopping through the spin flipping term, see Fig. 3). If the potential for up-spin electrons at site **j** is negative, the first way of hopping is energetically favorable. On the other hand, if the potential for up-spin electrons at site **j** is positive, then the second way of hopping becomes energetically favorable as shown in Fig. 3a. This means that in the off-site pairing state the electron can always hop along an energetically favorable path by flipping or not flipping spin according to the sign of the potential at site **j**. In other words, it means that the effective potential fluctuation $<V_{\mathbf{r}}^2>$ is largely reduced and therefore the potential scattering of quasiparticles is largely weakened in the off-site pairing state. For the on-site pairing state, however, only the first way of hopping is possible. In the case the potential for up-spin electrons at site **j** is positive, this hopping costs energy. The second way of hopping can be realized through the on-site spin flipping term, but it is not energetically favorable as the electron must first jump to a higher energy state either at site **i** (see Fig. 3b, hopping process 1) or at site **j** (Fig. 3b, hopping process 2). This shows that the enhancement of superconducting correlation to the mobility of electrons is weaker in the on-site pairing state than in the off-site pairing state in cases other conditions are same.

In 2D, the calculation becomes more complicated. Below we shall concentrate on a $d_{x^2-y^2}$-wave pairing state, $\Delta_\tau = \Delta(\delta_{\tau,\pm\mathbf{x}} - \delta_{\tau,\pm\mathbf{y}})$, whose energy gap vanishes at some points on the Fermi surface. To find the localization length in 2D, the localization length of quasiparticles on quasi-1D long strip with finite width M, i.e. $\lambda_M(E,V)$, is evaluated first. The bulk localization length of quasiparticles, $\lambda(E,V) = \lim_{M\to\infty} \lambda_M(E,V)$ is then determined from a single-parameter scaling ansatz

$$\frac{\lambda_M(E,V)}{M} = f(\frac{\lambda(E,V)}{M}). \quad (4)$$

In our calculation, M=7, 11, 15, 19, 23, 27, 31 are used. For the d-wave pairing state $\lambda_M$ grows extremely fast with increasing $\Delta$ or decreasing random potential V. To ensure that $\lambda_M$ converges fast enough to its infinite long strip limit, we limit the study only to small $\Delta$ pairing state with relatively large V.

The scaling function $f(x)$ and $\lambda(E,V)$ are determined from a least square fitting procedure [9]. To within numerical error, we find that the scaling function f(x) (nearly log(x) when x is sufficiently large) is the same for the d-wave state as for the normal state. This indicates that the d-wave pairing state belongs to the same universality class, i.e. the orthogonal class, as for the normal state, in agreement with the conclusion drawn based on the time-reversal symmetry consideration by Balatsky et al [11]. On the other hand, it implies that the conventional scaling theory is applicable and all quasiparticles are localized in the d-wave state in 2D.

Fig. ?? shows $\lambda_s/\lambda_n$ as a function of E for the d-wave pairing state. The behavior of $\lambda_s/\lambda_n$ for the d-wave state shows a great similarity to the 1D gapless pairing state in Fig. 1. But the enhancement of $\lambda_s/\lambda_n$ when E not too close to $E_F$ or to the band edge is much larger than its 1D counterpart. The enhancement of $\lambda_s/\lambda_n$ around the band edge is relatively small and nearly potential independent, similar to the 1D gapless states when they are far from the singularities of DOS (the DOS of a clean $d_{x^2-y^2}$-wave state diverges at an energy $E \sim \Delta$, not at the band edge). For the d-wave pairing state, the DOS at $E_F$ is largely suppressed by the opening of energy gap. But the corresponding $\lambda_s$ is not dramatically reduced compared with $\lambda_n$ at $E_F$. Like its 1D counterpart, $\lambda_s/\lambda_n$ at ($E_F$) for the d-wave pairing state is determined only by $\Delta$, independent of V. It decreases for small $\Delta$ and then increases for large $\Delta$ with a minimum located at around $\Delta \sim 0.25$, again similar to its 1D counterpart (Fig. 2).

The large enhancement of $\lambda_s/\lambda_n$ in nearly the whole energy band in Fig. ?? shows that the potential scattering of quasiparticles is largely weakened by the superconducting correlation and the localization effect is weaker in the d-wave pairing state than in the normal state. The sharp drop of $\lambda_s/\lambda_n$ in a narrow region around $E_F$ suggests that the states in this region are more localized than other low energy states. This agrees with the results of [1] and [3]. For an infinite layer system with each layer a d-wave superconductor, if we allow electrons to hop between layers, then those weakly localized states may first get delocalized. However, strongly localized states around $E_F$ may still be localized when the hopping constant between layers $t_\perp$ is sufficiently small. Hence a mobility edge may exist at low energy in this quasi-3D d-wave pairing state. The value of the mobility gap is $\Delta$ and V dependent, but it should be very small as Fig. ?? suggested.

In conclusion, we have studied the effect of superconducting correlation on the localization of quasiparticles in low dimensions. There is an intrinsic difference between an on-site pairing state (for example an isotropic s-wave pairing state) and an off-site pairing state (for example a d-wave pairing state) as far as the localization effect is concerned. The effect of superconducting correlation on the localization of quasiparticles is generally stronger in an off-site pairing state than in an on-site pairing state. $\lambda_s$ drops within the energy gap but is largely enhanced around an energy where a DOS peak shows. For the gapless pairing state in 1D and the d-wave pairing state in 2D, $\lambda_s/\lambda_n$ is enhanced in nearly the whole energy region and quasiparticles with energies close to $E_F$ are more strongly localized than other low energy quasiparticles but not more strongly than the corresponding normal states. For these gapless pairing states, $\lambda_s/\lambda_n$ at $E_F$ is determined only by the value of $\Delta$.

I wish thank J. M. Wheatley for many fruitful discussions.

[1] P. A. Lee, Phys. Rev. Lett. Phys. Rev. Lett. **71**, 1887 (1993).
[2] L. P. Gorkov and P. A. Kalugin, Pis'ma Zh. Eksp Teor. Fiz. **41**, 208 (1985) [JETP Lett. **41**, 253 (1985)]; K. Ueda and T. M. Rice, Theory of Heavy Fermions and Valence Fluctuations, edited by T. Kasuya (Springer, Berlin, 1985).
[3] Y. Hatsugai and P. A. Lee, Phys. Rev. B **48**, 4204 (1993).
[4] See, for example, D. Belitz, Phys. Rev. B **35**, 1636 (1987), and references therein.

[5] M. Ma and P. A. Lee, Phys. Rev. B **32**, 5658 (1985).
[6] C. L. Lambert and V. C. Hui, J. Phys.: Condens. Matter **2**, 7303 (1990); V. C. Hui and C. J. Lambert, *idib* **5**, 697 (1993).
[7] T. Xiang and J. M. Wheatley, Phys. Rev. B accepted.
[8] See, for example, E. Abrahams, P. W. Anderson, D. C. Licciardello, and T. V. Ramakrishnan, Phys. Rev. Lett. **42**, 673 (1979).
[9] A. MacKinnon and B. Kramer, Z. Phys. B **53**, 1 (1983), and references therein.
[10] Note that DOS becomes larger does not mean that $\lambda_s$ or $\lambda_n$ will also become larger. Actually, at the band edge where the maximum DOS shows, both $\lambda_s$ and $\lambda_n$ are very small (much smaller than at other energies). The large enhancement of $\lambda_s/\lambda_n$ around the DOS peak simply means that the effect of DOS on the localization length is stronger in superconducting states than in normal states.
[11] A. V. Balatsky, A. Rosengren, and B. L. Altshuler, Phys. Rev. Lett. **73**, 720 (1994).

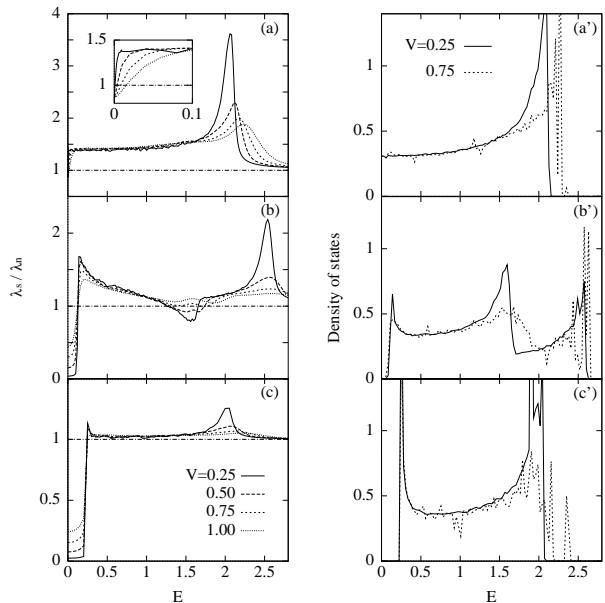

FIG. 1. (a) Ratio of the localization length between a superconducting state and the corresponding normal state $\lambda_s/\lambda_n$ vs energy E for four white noise potentials for a nearest neighbor pairing state $\Delta_\tau = \Delta\delta_{\tau=\pm 1}$ with $\mu = 0$ and $\Delta = 0.25$ in 1D. Inset: $\lambda_s/\lambda_s$ for small E. (b) Same as (a) but with $\mu = -0.5$. (c) Same as (a) but for an on-site pairing state $\Delta_\tau = \Delta\delta_{\tau=0}$. (a'), (b'), and (c') are density of states corresponding to (a), (b), and (c), respectively.

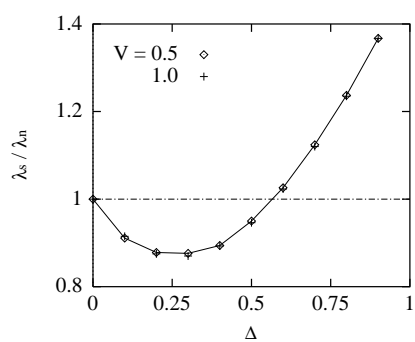

FIG. 2. $\lambda_s/\lambda_n$ vs energy gap $\Delta$ at $E_F$ for the off-site pairing state with $\mu = 0$ in 1D.

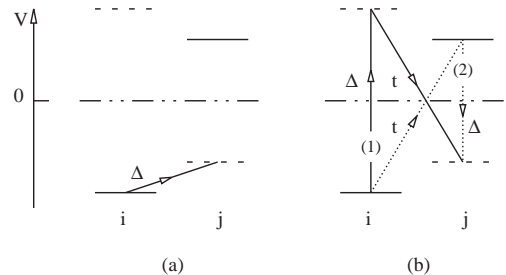

FIG. 3. Schematic representation of the pairing induced hopping processes of an electron from site **i** to site **j** in (a) an off-site pairing state and (b) an on-site pairing state. Scattering potentials are equal in amplitude but opposite in sign for up-spin and down-spin electrons at each site after a particle-hole transformation (see Hamiltonian $H'$), which are represented respectively by solid and dash horizontal lines.

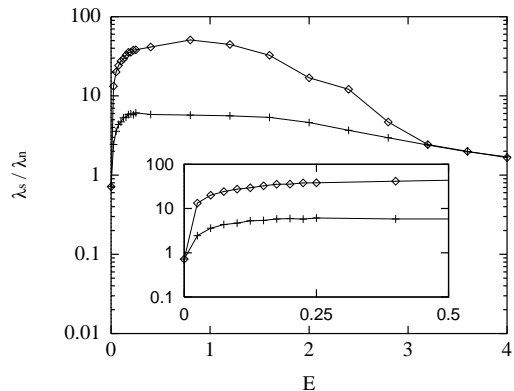

$\lambda_s/\lambda_n$ vs E for a d-wave pairing state with $\mu = 0$ in two white noise potentials, V=2 (diamond) and V=3 (cross), in 2D. Inset: $\lambda_s/\lambda_n$ for small E.

4